\documentclass[11pt]{article}

\usepackage{lipsum} 

\usepackage{bm}
\usepackage{floatrow} 
\pdfoutput=1
\usepackage{amsmath,amsfonts,amsthm}
\usepackage{esdiff}  
\usepackage{booktabs}  
\usepackage{url}  
\usepackage{hyperref}  

\usepackage[tableposition=top]{caption}

\usepackage{cleveref}  
	\crefname{equation}{equation}{equations}
	\crefname{figure}{figure}{figures}	
	\crefname{table}{table}{tables}
\usepackage[skip=1.5pt,font=small]{caption}

\usepackage{bbm}

\usepackage[usenames,dvipsnames,svgnames,table]{xcolor}

\usepackage{graphicx}

\usepackage{tikz}  

\usepackage[sc]{mathpazo} 
\usepackage[T1]{fontenc} 
\usepackage{microtype} 

\usepackage{multicol} 
\usepackage[margin={1cm,1.5cm}]{geometry}
\usepackage{booktabs} 
\usepackage{float} 
\usepackage{hyperref} 

\usepackage{lettrine} 
\usepackage{paralist} 

\usepackage{abstract} 

\usepackage{titlesec} 
\renewcommand\thesection{\Roman{section}} 
\renewcommand\thesubsection{\Alph{subsection}} 
\titleformat{\section}[block]{\large\scshape\centering\bfseries}{\thesection.}{1em}{} 

\titleformat{\subsection}[block]{\scshape\centering}{\thesubsection.}{1em}{} 

\usepackage{fancyhdr} 
\DeclareCaptionFormat{myformat}{#1#2#3\hrulefill}
\usepackage{float}
\floatstyle{plaintop}
\restylefloat{table}

\usepackage{authblk}
\makeatletter
\title{\vspace{-15mm}\fontsize{16pt}{16pt}\selectfont\textbf{Forecasting Future Murders of Mr. Boddy by Numerical Weather Prediction}} %

\author[1,2]{Eve Armstrong\thanks{aeve@sas.upenn.edu}}
\affil[1]{Computational Neuroscience Initiative, University of Pennsylvania, Philadelphia, PA 19104, USA}\par
\affil[2]{Department of Physics, New York Institute of Technology, New York, NY 10023, USA}
\date{(Dated: April 1, 2019)}
\begin{document}
\maketitle 

\begin{abstract}
\noindent
Despite a previous description of his state as a stable fixed point, just past midnight this morning Mr. Boddy was murdered again.  In fact, over 70 years Mr. Boddy has been reported murdered $10^6$ times, while there exist no documented attempts at intervention.  Using variational data assimilation, we train a model of Mr. Boddy's dynamics on the time series of observed murders, to forecast future murders.  The parameters to be estimated include instrument, location, and murderer.  We find that a successful estimation requires three additional elements.  First, to minimize the effects of selection bias, generous ranges are placed on parameter searches, permitting values such as the Cliff, the Poisoned Apple, and the Wife.  Second, motive, which was not considered relevant to previous murders, is added as a parameter.  Third, Mr. Boddy's little-known asthmatic condition is considered as an alternative cause of death. 

Following this morning's event, the next local murder is forecast for 17:19:03 EDT this afternoon, with a standard deviation of seven hours, at \textit{The Kitchen} at 4330 Katonah Avenue, Bronx, NY, 10470, with either the Lead Pipe or the Lead Bust of Washington Irving.  The motive is: Case of Mistaken Identity, and there was no convergence upon a murderer.  Testing of the procedure's predictive power will involve catching the D train to $205^{th}$ Street and a few transfers over to Katonah Avenue, and sitting around waiting with our eyes peeled.  We discuss the problem of identifying a global solution - that is, the best reason for murder on a landscape riddled with pretty-decent reasons.  We also discuss the procedure's assumption of Gaussian-distributed errors, which will under-predict rare events.  This under-representation of highly improbable events may be offset by the fact that the training data, after all, consists of multiple murders of a single person.
\end{abstract}
\section{INTRODUCTION}
\begin{multicols}{2}
When we last visited Mr. Boddy Sylvester van Meersbergen, he lay motionless on the concrete floor of an aviary in Philadelphia, Pennsylvania, just west of the Schuylkill River~\cite{armstrong2018colonel}.  His dynamics had transitioned from a limit cycle, via an Andronov-Hopf bifurcation that has since been attributed to Colonel Mustard with the Candlestick.  The medical authorities present at the time concurred that his configuration represented an asymptotically stable equilibrium.  Nevertheless, within mere minutes Mr. Boddy was murdered again~\cite{murderAfter2018}.

In fact, in the interim year Mr. Boddy has been reported murdered 14,322 times~\cite{hasbro}, most recently just this morning~\cite{murderThisMorning}.  Moreover, there exists a 70-year time series of such reports~\cite{clueGame}, with a mean frequency of $10^{-3}$ Hz.  To our knowledge, no attempt has been made at employing this record of observations for predictive purposes, with the aim of intervention or even just for sport.  Yet a method of prediction is readily at hand.  In the field of numerical weather prediction, it is called data assimilation.

Data assimilation (DA) was invented to predict the weather~\cite{betts2010practical,evensen2009data,kalnay2003atmospheric,kimura2002numerical}.  It is a means to exploit information in observable data to complete a dynamical model of the physical system that generated the data.  This model may have unknown parameters to be estimated, and may include state variables that are not observed.  If the transfer of information from measured to unmeasured quantities is efficient, the completed model can be integrated forward to predict future data.  DA differs from the formulation of machine learning in that it assumes that the data were generated by a physical system rather than sorcery~\cite{honegger2018shedding}. 

The DA framework is apt for capturing the case of Mr. Boddy.  Mr. Boddy is a known dynamical system, who at any time is observable as either Alive or Dead\footnote{Quantum considerations are beyond the scope of this paper; see \textit{Discussion}.}.  His trajectory in state space can be described in terms of Markov transition probabilities, which, importantly, capture his ability to instantly forget having died.  

Multiple external parameters act upon Mr. Boddy, including instrument, location, and murderer, and at any timepoint on his trajectory each parameter takes a specific value.  We would like to use the record of reported deaths to estimate those parameters at each instance on the trajectory, to predict future instances of death.  The 70-year time series of observations is significantly longer than the temporal distance between murders.  Thus we have a decent shot at using that record to predict a small distance into the future.

We find that several additional ingredients are required to converge on a solution.  First we need motive, or more generally, \lq\lq reason\rq\rq, as a fourth parameter governing Mr. Boddy's dynamics.  Second, we ameliorate the selection bias introduced by using information only from \textit{reported} deaths.  Specifically, we place generous limits on the search ranges of all parameter values, tossing in extremely-likely (Figure 1) and extremely-unlikely (Figure 2) values that have never been reported.  For example, Undercooked Chicken (likely) and the Poisoned Apple (unlikely) are included as instruments.  We also allow each candidate murderer to disguise him-or-herself as any other candidate murderer, so that a serial killer may be pulling off the entire time series single-handedly\footnote{\lq\lq Serial killer\rq\rq\ refers to multiple murders by one person.  We permit a special case of multiple murders by one person \textit{of} one person.}.

Third, we consider Mr. Boddy's little-known severe asthmic condition: occasionally, Mr. Boddy becomes fatally allergic to breathing.  His general practitioner attests that it is as likely a cause of death as is murder~\cite{generalPractitioner}, even given the observed murder frequency.  Thus, we add as an unobserved state variable Mr. Boddy's level of Immunoglobulin E (IgE) antibodies.  Heightened levels of IgE antibodies indicate an overreaction to allergens, and are associated with fatality in some autopsies~\cite{sur1994sudden}.

The DA procedure used in this paper is an optimization wherein a cost function is minimized.  We discuss details of the procedure, including the problem of iden-
\end{multicols}
\begin{figure}[H]
\centering
  \includegraphics[width=0.6\textwidth]{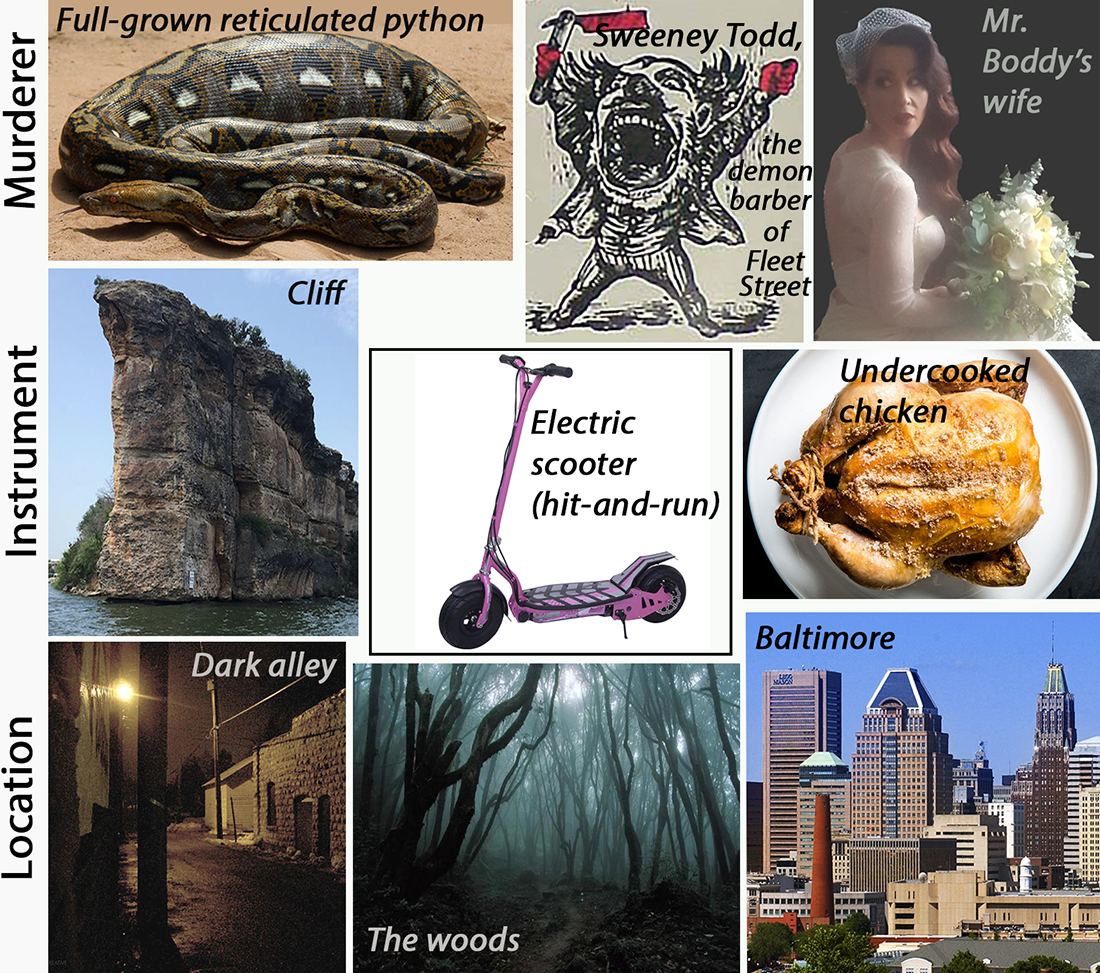}
\vspace{3mm} 
   \caption{\textbf{Highly likely values of $\theta_{murder}$ that may have been historically undersampled.}  (Left-to-right, top-to-bottom: \cite{python,sweeney,sara,cliff,scooter,chicken,darkAlley,woods,baltimore}.)}
\end{figure}
\begin{figure}[H]
\centering
  \includegraphics[width=0.6\textwidth]{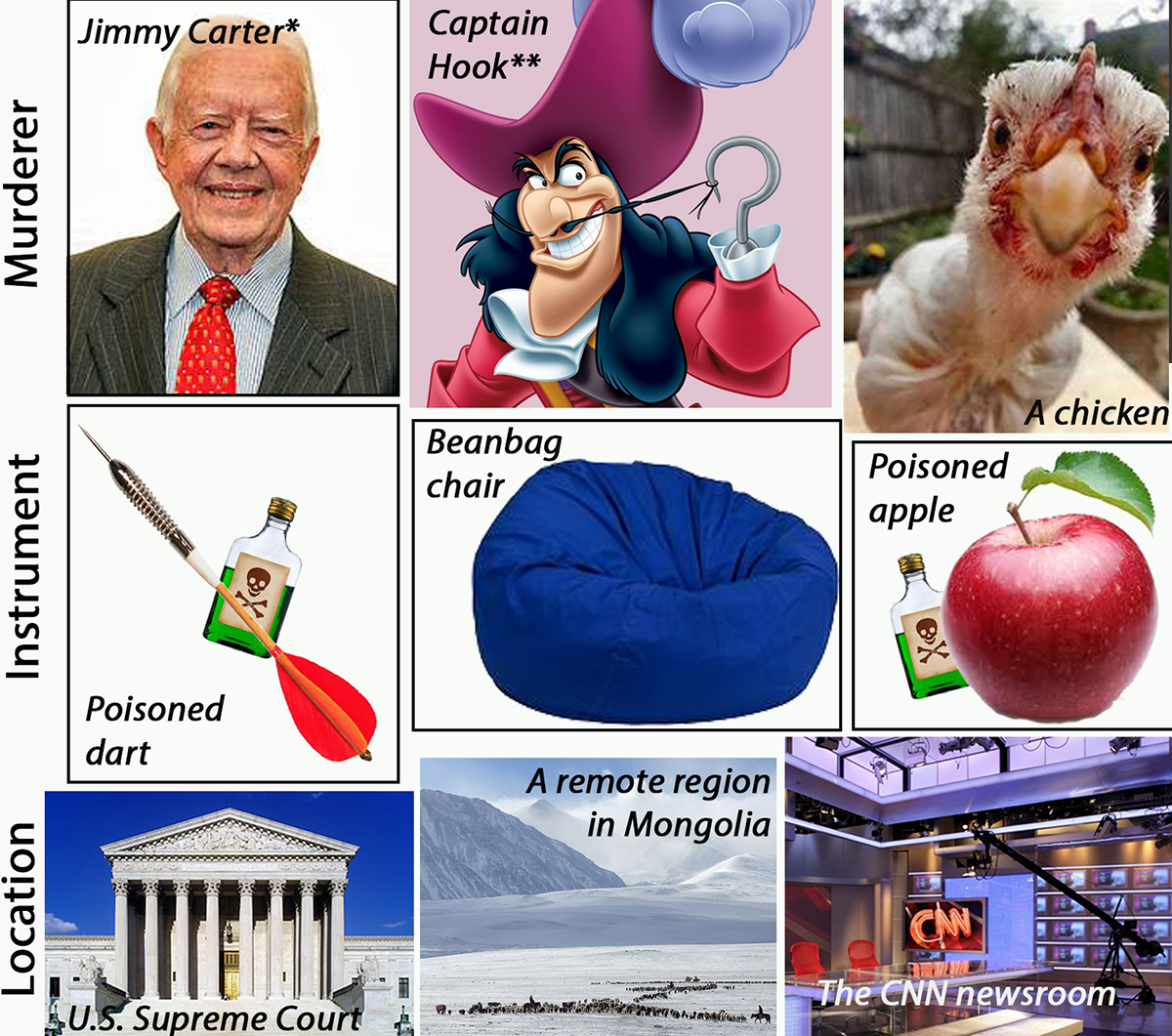}
\vspace{3mm} 
   \caption{\textbf{Highly unlikely values of $\theta_{murder}$, tossed into the parameter search ranges for good measure.}  *Too nice.  **Too obvious.  (Left-to-right, top-to-bottom: \cite{jimmyCarter,captainHook,aliveChicken,dart,beanbagChair,apple,supremeCourt,mongolia,CNN}.)}
\end{figure}
\begin{multicols}{2}
\noindent
tifying a minimum that is low enough to satisfy our predictive purposes.  Our predictive purposes are not yet clear to us, and so at the moment this is not a high-priority concern.  The procedure also assumes Gaussian errors, which will under-predict rare events.  This is troublesome, as rare events - when they do occur - are typically the most unsettling.  And, at least in the case of the weather, the most deadly.

The first murder forecast within the local region is late this afternoon.  We are reasonably confident that this will happen, as a preliminary DA prediction that used an 80-versus-20\% (trial-versus-test) split of the 70-year time series reached one Lyapunov time in one week.  That is, we won't reach the limit of predictability until April 8.  Well, only time will tell.  Now we wait.
\end{multicols}
\section{MODEL}
\begin{multicols}{2}
In this Section we define Mr. Boddy's dynamical map.  Let a model $\bm{f}: \bm{X}_n \to \bm{X}_{n+1}$ define a forward mapping of a dynamical system $\bm{X}$.  We assume Markov chain transition probabilities~\cite{privault2013understanding}, where the state of the system at time ($n+1$) is entirely determined by the state at time $n$.  This is an important feature for Mr. Boddy, who evidently forgets having been murdered just a timestep ago.  The model is a function of dynamical variables $\bm{x}$ and of any unknown model parameters $\bm{\theta}$, where the parameters themselves may be time-varying.  That is, $\bm{X} \equiv [\bm{x}_{t=0},\bm{x}_{t=1},...,\bm{x}_{t=T},\bm{\theta}_{t=0},\bm{\theta}_{t=1},...,\bm{\theta}_{t=T}]$, where $T$ is the final time step.  The parameters are distinguished from the state variables because, while we write a forward mapping for the physical system under study, the parameters are external forces whose underlying dynamics are unknown.

For the purposes of this study, the only important characteristic of Mr. Boddy is whether he is alive or dead.  This is a directly observable quantity with a 70-year record of reported values (Figure 3), and we define it as a state variable.  Now, in initial tests of the DA procedure, when we wrote Mr. Boddy's state solely in terms of this one state variable, there was no convergence on a solution.  For this reason, we consulted Mr. Boddy's general practitioner to ask whether we were missing a piece of the picture.  It turns out that in addition to his
\end{multicols}
\begin{figure}[H]
\centering
  \includegraphics[width=0.9\textwidth]{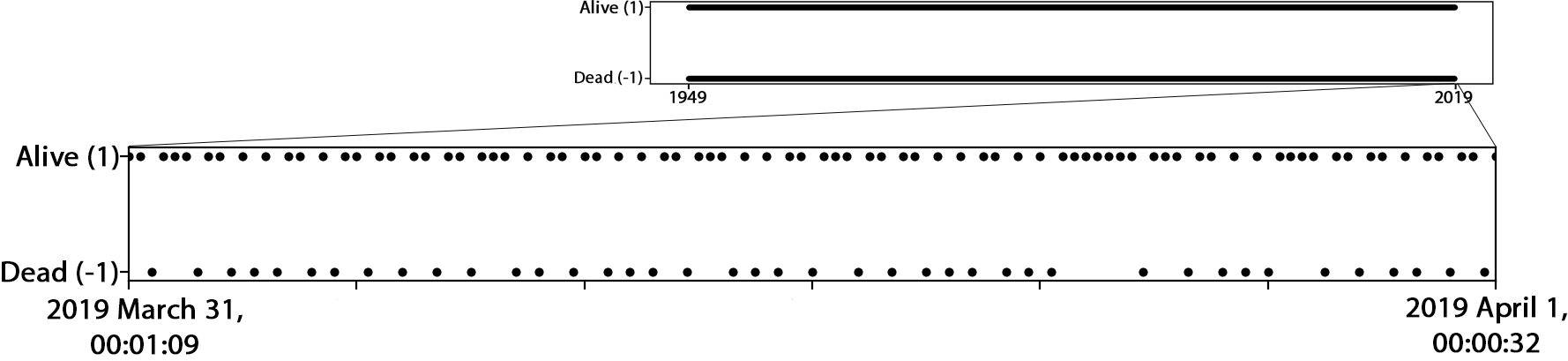}
\vspace{3mm} 
  \caption{\textbf{Time series of observations, modeled as Markov-chain transitions from Alive to Dead.}  \textit{Top}: Full 70-year time series.  The jumps cannot be resolved by-eye.  \textit{Bottom}: Zoom-in on the final 23 hours.  Alive-to-dead transitions are governed by the transition probabilities of Equation Set 2; dead-to-dead transitions do not occur.}
\end{figure}
\begin{multicols}{2}
\noindent
 proclivity to being murdered, there exists another likely means for Mr. Boddy to transition to Dead: a severe asthmatic condition (see \textit{Introduction} for references).  We thus added an unmeasured state variable: Mr. Boddy's level of IgE antibodies\footnote{The reader may wonder why the state variable representing asthmatic condition is defined to be unmeasurable, given that there exist measurable quantities used to monitor asthma.  In fact, Mr. Boddy does not use any of them.  This may seem reckless, but given how sportingly he weathers being murdered, perhaps his disregard for modern medicine is understandable.}.  
 
The dynamical system is defined, then, in terms of two state variables: $\bm{x} \equiv  [x,y]$, where $x$ is the Alive-versus-Dead state and $y$ is the level of IgE antibodies.  The parameters governing murder ($\bm{\theta}_{murder} \equiv$ [\textit{instrument, location, murderer, reason}]) are distinct from those governing asthmatic condition ($\bm{\theta}_{asthma} \equiv$ [\textit{concentration of airborne Alternaria\footnote{Alternaria is a genus of fungi that can cause allergies in humans~\cite{bush2004alternaria}.}, and other environmental factors}\footnote{We include in $\theta_{asthma}$ other environmental factors in addition to Alternaria, which are described in a twenty-seven page pamphlet that Mr. Boddy's general practitioner gave us.  Upon request, the reader is welcome to peruse it.}]).  Note also the addition of reason as a parameter, which is required for solutions to converge.  Some reasons are listed in Table 1.

We now write the continuous-time dynamics most generally as:
\begin{equation}
\begin{aligned}
  &\diff{x}{t} = f_x(x(t), y(t,y_{thresh}), \bm{\theta}_{murder}(t));\\
  &\diff{y}{t} = f_y(x(t), y(t,y_{thresh}),\bm{\theta}_{asthma}(t)),
\end{aligned}
\end{equation}
\noindent
where the affine parameterization $t$ is time, $y$ is continuous, and $x$ may take discrete binary values of $1$ (Alive) or $-1$ (Dead)\footnote{Dead is defined as $-1$ to make the imposition of unitarity in the cost function computationally simpler; see \textit{Estimation and Prediction}.}.  The number $y_{thresh}$ is the value of IgE antibody level above which Mr. Boddy will die of an asthma attack.  

To write a specific form for Equation Set 1, let us examine the coupling of $x$ and $y$.  If the value of $y$ at time $n$ (denoted $y_n$) is above $y_{thresh}$, then $x_{n+1}$, by definition of $y_{thresh}$, will be $-1$; that is, Mr. Boddy will die from an asthma attack.  On the flip side, if $x_n = -1$ already, then $y$ will not evolve in the interim before time ($n+1$), as Mr. Boddy's biological processes have momentarily ceased to run.  Neither of these cases is particularly interesting.  

A more tantalizing scenario presents itself when $x_n = 1$ and $y_n < y_{thresh}$.  At such a time, Mr. Boddy is alive and will not die from asthma within the next time step - and thus we can focus on murder probability.  Let's examine this simplified (i.e. no-asthma-attack) regime, focusing on variable $x$, and recouching the equations in terms of probabilities.  We write:
\begin{equation}
\begin{aligned}
  &P(x_{n+1}=-1|x_n=1,y_n < y_{thresh}) = f_x(\bm{\theta}_{murder_n});\\
  &P(x_{n+1}=-1|x_n=-1,y_n < y_{thresh}) = 0.
\end{aligned}
\end{equation}
The first line in Equation Set 2 states that the probability of transitioning from Alive to Dead, given $y_n < y_{thresh}$, is some function of the parameters $\bm{\theta}_{murder}$.  The second line states that the probability of transitioning from Dead to Dead is zero (a step function will be employed to transition him back to Alive).  In other words, the probability rules are: If Alive and murdered, transition to Dead; if Dead, transition to Alive.

Let us expand the most interesting case of Equation 2, Line 1:
\end{multicols}
\setlength{\tabcolsep}{1pt}
\begin{table}[htb]
\small
\centering
\caption{\textbf{Some Possible Reasons Why Mr. Boddy Keeps Getting Murdered}} 
\begin{tabular}{ l c} \toprule
 \textit{He was mean.}  \\
 \textit{He was devastatingly handsome (or some trait that inspires envy/spite).} \\
 \textit{He was burglarized.}\\
 \textit{He was blackmailing the murderer for a large sum of money over a secret that the murderer does not want divulged.} \\
 \textit{He witnessed a different murder and needed to be silenced.}\\
 \textit{They were cases of mistaken identity.}  \\ 
 \textit{They were accidents.} \\ 
 \textit{He was not murdered; he died naturally from his pre-existing asthmatic condition.}\\ \\\bottomrule
\end{tabular}
\newline
\end{table}
\begin{equation}
\begin{aligned}
  P(x_{n+1}=-1) =\, &f_x(\bm{\theta}_{murder}) \\
=\, & f_x(P(instrument|location,P(instrument|murderer,P(instrument|reason,\\
&P(murderer|location,P(murderer|reason,P(reason|location))))))).\\ \vspace{0.1cm}
\end{aligned}
\end{equation}
\begin{multicols}{2}
\noindent
Equation 3 reads: \textit{The probability that Mr. Boddy dies is a function of the probability of instrument conditioned on location, conditioned on the probability of instrument conditioned on murderer, conditioned on the probability of instrument conditioned on reason, conditioned on the probability of murderer conditioned on location, conditioned on the probability of murderer conditioned on reason, conditioned on the probability of reason conditioned on location}.  No causality is assumed, and thus the ordering of these conditional probabilities is arbitrary.  Importantly, while the parameters $\bm{\theta}_{murder}$ may in general be independent, they are \textit{not} independent when a successful murder is pulled off.  For example, if the instrument is a Mack Truck and the location is a dining room, then a murder is unlikely to occur, because it is nontrivial to squeeze a Mack Truck into most dining rooms.  If a murder occurs and the murderer is blind, the instrument is unlikely to be a Mack Truck.  Also, for a murder to occur the instrument and murderer must appear simultaneously at one location.  

To perform calculations, each distribution $P(m|n)$ of Equation 3 is written as an $m \times n \times (T+1)$ matrix of numbers in the range [0:1), where $T$ is the number of time steps.  These values are calculated directly from the 70-year record.  For the new parameter values that we added ourselves, we just assumed some statistics.  We are unable to justify this assumption, but shall wait through the prediction window to see whether it did any harm. 

Finally, we examine the evolution of $y$, the unmeasured level of IgE antibodies.  Taking the case where Mr. Boddy is alive at time $n$, $y_{n+1}$ is governed by $y_n$, $y_{thresh}$, and $\bm{\theta_{asthma}}$.  If $y_n > y_{thresh}$, then at time $(n+1)$ Mr. Boddy dies and $y$ will be reset to Mr. Boddy's baseline value.  Otherwise $y_{n+1}$ will depend on $f_y(\bm{\theta_{asthma}})$.  The form of $f_y(\bm{\theta_{asthma}})$ is explained in a 27-page medical pamphlet given to us by Mr. Boddy's general practitioner, which is available upon request. 
\end{multicols}
\section{ESTIMATION AND PREDICTION}
\begin{multicols}{2}
\subsection{\textbf{The problem}}

Armed with observations and a model, we proceed to estimation and prediction, using data assimilation.  DA is an inverse formulation, whereby information contained in observations is harnessed to complete a model of the system from which the observations were obtained, thereby informing us of where the information came from~\cite{tarantola2005inverse}.  In case that sounds like circular logic: it is not.   

Generally, the model is written as:
\begin{align*}
  \diff{x_a(t)}{t} &= f_a(\bm{x}(t),\bm{\theta}); \hspace{1em} a =1,2,\ldots,D,
\end{align*}
\noindent
which says what Equation 1 says for $D = 2$.  Aside from the unknown parameters, the model is assumed to be correct.  In case that sounds foolish: it might be.

We seek to ascertain whether sufficient information can be propagated through the coupled equations during the window in which we have observations (the estimation window) to estimate the parameters and the dynamics of the unmeasured variables.  The test of a successful estimation is its ability to predict the model's behavior in regions outside the window of observations - for example, the future\footnote{Predicting the past or the middle can also be fair game.}.

\subsection{\textbf{The most likely hypothesis}}

In this paper we use optimization to solve this inverse problem.  To scare up something to optimize, we define a discretized state space in which our model lives.  Within the estimation window, this space is a $(D + N_{\theta})(T+1)$-dimensional lattice, where $N_{\theta}$ is the number of parameters and $T$ is the number of time steps.  With $D=2$, $N_{\theta} = N_{\theta_{murder}} + N_{\theta_{asthma}} = 88 + 134$, and $T = 2,207,520,001$ for a step size of one second, the state space is 494,484,480,224-dimensional.  

The series of lattice points that the model traverses during the estimation window is a path $\bm{X}$.  That is, for our model, $\bm{X} = [\bm{x}(t),\theta_{murder}(t),\theta_{asthma}(t)] = [x(t),y(t),\theta_{murder}(t),\theta_{asthma}(t)]$.   The path $\bm{X}$ defines the set of estimated values of all variables and parameters at each timepoint that produced the observed deaths of Mr. Boddy within the 70-year history of observations.  

Now, some of the reported deaths include information regarding instrument, location, and/or murderer, while some are incomplete.  No reports include information on reason or variable $y$.  Not all events were necessarily reported, and the time step between reports is variable and typically larger than the time step of the dynamical model evolution.  In short, there exists much uncertainty - and consequently much wiggle room in defining a path that explains the data.  In fact, there exists an infinite number of possible paths.  We seek the path that is the one \textit{most likely} to have produced the observations.  This most likely path is denoted $\bm{X}_0$.  

How do we find path $\bm{X}_0$?  The 494,484,480,224-dimensional state space in which it lives is rather voluminous, and it would be a formidable task to define the geometry of that space.  Fortunately, we do not need to know the shape of the space itself.  We can define $\bm{X}_0$ in terms of the observations and the dynamics.  

Let us first consider how the observations, denoted $\bm{S} = [s_{t=0},s_{t=1},...,s_{t=T}]$\footnote{Generally, the observations $\bm{S}$ consist of a set of $\bm{s}_l$ distinct measured quantities.  In our case we only observe one quantity: whether Mr. Boddy is alive or dead.  Thus, $l=1$ and $\bm{S}$ can be simplified as written above.}, define $\bm{X}$.  If we write:
\begin{equation}
  P(\bm{X}|\bm{S}) = e^{-C(\bm{X},\bm{S})},
\end{equation}
then the most probable path $\bm{X}$ given observations $\bm{S}$ is the path that minimizes some quantity $C$.  $C$ we will treat as a cost function to be minimized.  

To calculate $\bm{X}_0$\footnote{In addition to calculating the expectation value of $\bm{X}$ itself, we can calculate the expectation value of any known function $G(\bm{X})$ as: $\langle G(\bm{X}) \rangle = \frac{\int d\bm{X} G(\bm{X}) e^{-C(\bm{X},\bm{S})}}{\int d\bm{X} e^{-C(\bm{X},\bm{S})}}$.  That is, $G(\bm{X})$ is written as a weighted sum over all possible paths, with weights exponentially sensitive to $C$.}, we need a useable form for $C$.  Toward this end, let us rearrange and rewrite Equation 4, and for the moment focus on the observations (ignoring the model):
\begin{equation}
\begin{aligned}
  C(\bm{X},\bm{S}) &= - \log P(\bm{X}|\bm{S})\\
                             &= - \sum_n \text{CMI}(\bm{x}(n),\bm{s}(n)|\bm{s}(n-1))\\
                             &= - \sum_n \text{MI}(\bm{x}(n)|\bm{s}(n)).
\end{aligned}
\end{equation}  
\noindent
On the second line of Equation Set 5, $P(\bm{X}|\bm{S})$ is defined in terms of the conditional mutual information (CMI)~\cite{wyner1978definition}: the probability of the current state conditioned on the current observation, conditioned on all previous observations.  Next we assume that the observations at different times are independent of each other, which is a ridiculous assumption, and which yields the simplified third line, involving the mutual information (MI)~\cite{cover2012elements} between the state and the observation at time $n$. 

Now let's bring the model dynamics into the scheme.  The Markov transition probabilities may contain errors, and we add a second term to allow for that:
\begin{align*}
  C(\bm{X},\bm{S}) = - \sum \text{MI}(\bm{x}(n)|\bm{s}(n)) - \sum \log[P(\bm{X}_{n+1}|\bm{X}_n]. 
\end{align*}  
Errors in the guess of initial conditions of the path should also be considered, which would add a term $-\log[P(\bm{x}_0)]$.  We, however, do not know the initial conditions.  Thus we shall assume that the minimizing path is independent of initial conditions, and wait to see how egregiously this assumption impacts the prediction.  

Toward a useable form of $C$, we make two more ridiculous assumptions.  First, both measurement and model errors have Gaussian distributions.  That is, highly improbable events probably will never happen (see \textit{Discussion}).  Second, there is no transfer function between Mr. Boddy's measured Alive-versus-Dead state versus his modeled Alive-versus-Dead state.  Or: all accounts of occurrences of Dead are infinitely accurate and beyond question.  Finally, we write:
\end{multicols}
\begin{align}
  C &= \overbrace{\sum_{j} \frac{A}{2}(s_l(n) - x_{a=1}(n))^2}^{error_{measurement}} + \overbrace{\sum_{n}^{N-1}\sum_{a}^{D} \frac{B^{a}}{2}\left(x_a(n+1) -
f_a(\bm{x}(n),\bm{\theta})\right)^2}^{error_{model}} + \overbrace{\lambda\sum_n^{N}(|x_{a=1}(n)|-1)^2}^{error_{unitarity}},
\end{align}
\begin{multicols}{2}
\noindent
where $t_{n} \equiv n$, \textbf{$s_l$} are measurements (or observations), \textbf{$x_a$} are state variables, and $\bm{\theta}$ are unknown parameters.  The coefficients $A$ and $B$ are inverse covariance matrices for the measurement and model errors, respectively.  In the first term, a least-squares measurement error has been boiled down from mutual information.  Only the variable $x$ is measured, and success of the estimation will required an efficient transfer of information from the measured state $x$ to unmeasured state $y$, as we require a completed model in order to generate predictions.  

The second term of Equation 6 permits error in the model evolution - for example, stemming from the fact that the discretized state space does not have infinite resolution.  The third term is an equality constraint added to impose unitarity: Mr. Boddy is never created or destroyed, but rather is either alive or dead (but see \textit{Discussion} on quantum effects).  For a thorough derivation leading from Equation 4 to 6, see Ref \cite{abarbanel2013predicting}.   

\subsection{\textbf{Estimation}}

The estimate of path $\bm{X}_0$ will consist of $(D + N_{\theta})(T+1) = 494,484,480,224$ numbers.  The search of the state space for these numbers is performed via the variational method~\cite{smith1998variational}, requiring that the first derivative of $C$ with respect to the minimizing path be zero, and that its second derivate be positive-definite.  The variational method is descent-only, which presumes that the correct search direction is always Downward.  To avoid presuming this, one may opt for Monte Carlos~\cite{wong2019precision} instead. 

If there exist degenerate solutions - for example, multiple explanations for a single crime - then finding the minimizing path is nontrivial (Figure 4).  To identify a global minimum\footnote{Actually, we don't necessarily have to find the deepest minimum; we just need a minimum that is deep enough for our predictive purposes.  As we have yet to define our predictive purposes, perhaps the dart-throwing is premature.} amid a sea of shallower minima, we employ a method of iterative dart-throwing~\cite{van1987simulated}.

\subsection{\textbf{Prediction}}

Prediction involves taking as initial conditions the values of the variables and parameters\footnote{The discriminating reader will wonder how we plan to integrate forward time-varying parameters with unknown dynamics.  We have no plan, and opt to ignore the problem.  Thus we hope that the reader is not discriminating, and - by this Page 8 - has tired of reading footnotes.} at the final time step of the estimation window, and integrating.  The full DA routine will be continually updated and amended as new deaths are reported, to strengthen forecast accuracy.  

\subsection{\textbf{The parameter search ranges and rare events}}

As noted, taking as observations the 70-year reported history invites selection bias into the estimation procedure.  Some reported deaths were poorly described, possibly because the investigations were not seen to completion.  Further, not necessarily all deaths were reported.  In addition, even the complete reports appear to draw the instrument, location, and murderer from rather rigid boundary conditions.  The location options, for example, are nine rooms in a single house.  This might be due to laziness on the part of the investigators, who didn't feel
\begin{figure}[H]
\centering
  \includegraphics[width=65mm]{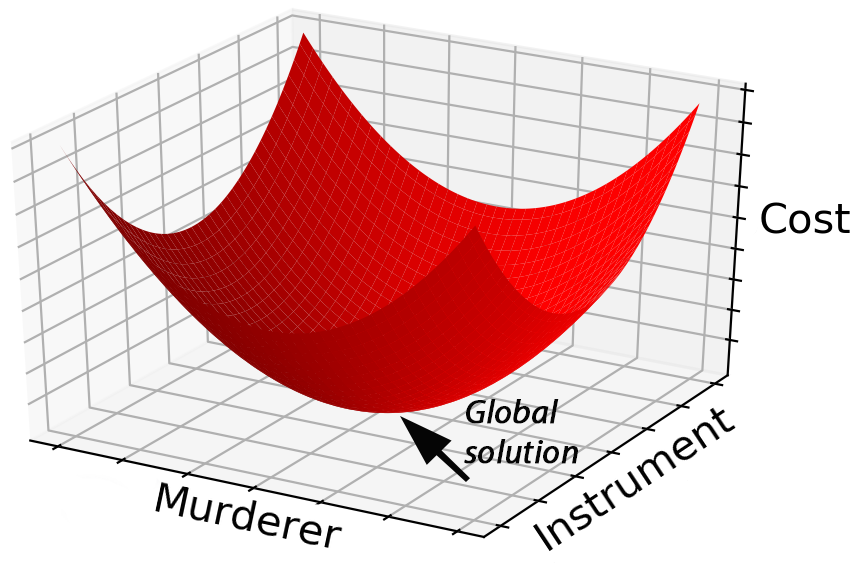}\\
  \includegraphics[width=65mm]{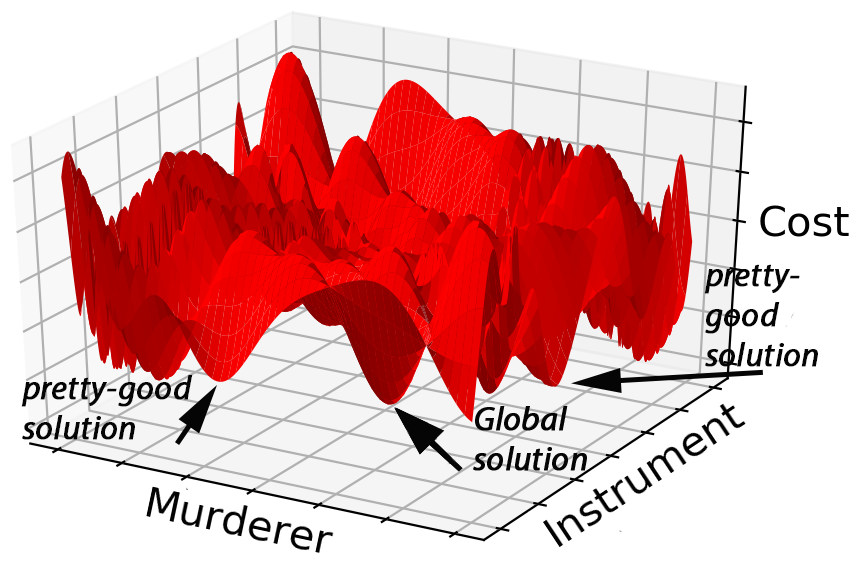}
\vspace{3mm} 
  \caption{\textbf{The surface of the cost function $C$: a three-dimensional representation of the 494,484,480,224-dimensional state space, with example directions \textit{murderer} and \textit{instrument}.}  \textit{Top}: The cost function is smooth and convex, and thus there exists one global minimum corresponding to the most likely path $\bm{X}_0$.  \textit{Bottom}: A more realistic scenario~\cite{eggholderFunction} with multiple possibilities that may muddy the investigation.}  
\end{figure}
\end{multicols}
\setlength{\tabcolsep}{1pt}
\begin{table}[htb]
\small
\centering
\caption{\textbf{Red Flags: Overestimates of Rare Events}} 
\begin{tabular}{ l c} \toprule
 \textit{It's a serial killer.} \\
 \textit{It's a serial killer with a different motive each time.} \\
 \textit{All murders were cases of mistaken identity*.}\\
 \textit{All murders were accidents.}\\
 \textit{Unlikely pairings of parameters occur, e.g. Dining Room and Mack Truck.}  \\
 \textit{Any of the highly-unlikely parameter values (Figure 2) is overrepresented.} \\ \\\bottomrule
\end{tabular}
\newline
\small{As the assumption of Gaussian-distributed errors will under-predict rare events, the overestimation of any of the above events will be cause for concern.  *Actually, we are not certain that the mistaken murder of Mr. Boddy $10^6$ times is necessarily less probable than the intentional murder of Mr. Boddy $10^6$ times.}
\end{table}
\begin{multicols}{2}
\noindent
like extending their detective work outdoors.  Indeed, some of what are typically considered highly likely locations, such as the Dark Alley or The Woods, may have been undersampled.  Moreover, we found that when our search ranges included only the reported parameter values, estimates asymptoted to the bounds - a sign that the bounds needed extending.

We extended the permitted values of all model parameters to include both highly-likely and highly-unlikely values (see Figures 1 and 2, respectively, for examples).  In addition, we added a few ludicrously-unlikely values, such as the Asteroid, as a check on the DA procedure.  The assumption of Gaussian errors should under-predict rare events, so if a ludicrously-unlikely event pops up frequently in the estimation window, it is cause for concern.  Finally, some \textit{combinations} of parameter values are particularly unlikely.  For a list of some of these \rq\rq red flags\rq\rq, see Table 2.
\end{multicols}
\section{RESULT}
\begin{multicols}{2}
Preliminary tests were performed using 80\% of the 70-year time series as training data, with the remaining 20\% reserved for prediction.  These tests confirmed that motive and asthma indeed are required for convergence.  They also indicated that the search ranges required expanding beyond their historically-defined values, as otherwise the estimates asymptoted to the bounds.   These preliminary tests marked one Lyapunov time\footnote{The Lyapunov time defines the characteristic timescale on which the dynamics of a system can be predicted (e.g. Ref~\cite{pikovsky2016lyapunov}).} at roughly a week.  That is, the  procedure stands a decent chance at predicting deaths through April 8. 

As noted, we expanded the parameter search ranges to include ludicrously-unlikely values.  In addition to providing a confidence check on the estimations, this strategy was also intended to help distinguish between an historically unreported value because it wasn't looked for, versus an unreported value because it is intrinsically rare.  Unfortunately, expanding the search ranges to this ludicrous extent increased the computational expense such that convergence failed.  Thus, unfortunately we were unable to probe the likelihoods of what arguably are the most entertaining possibilities.  Our assumption of Gaussian errors may have precluded such findings anyway, although that assumption may in part be offset by the fact that the procedure was trained upon an entire time series of extremely rare events, namely: multiple murders of one person.

The best estimations were those that omitted the ludicrously-unlikely parameter values.  Estimations for location, for example, are shown in Figure 5.  Here, red and blue denote regions of correctly-estimated reported deaths, and erroneous estimations, respectively.  The estimations over-all appear reasonable, and common locations pop up frequently.  There are a few peculiarities, including erroneous estimations throughout the state of Connecticut and on small islands near the Arctic circle.  In addition, we were surprised to note dense - and correct - estimations in Siberia. 

The prediction window for state variable $x$ is shown in Figure 6.  For the Dead instances, circles and triangles represent murders and asthma attacks, respectively.  The first predicted regional event is: 17:19:03 EDT today, with a standard deviation of seven hours, at \textit{The Kitchen} at 4330 Katonah Avenue, Bronx, NY, 10470 (Figure 7), with either the Lead Pipe or the Lead Bust of Washington Irving.  The motive is: Case of Mistaken Identity; there was no convergence in the state space direction of murderer.  The location is pinpointed quite accurately, which simplifies the next step: to test the result, we know where to go.
\end{multicols}
\begin{figure}[H]
\centering
  \includegraphics[width=160mm]{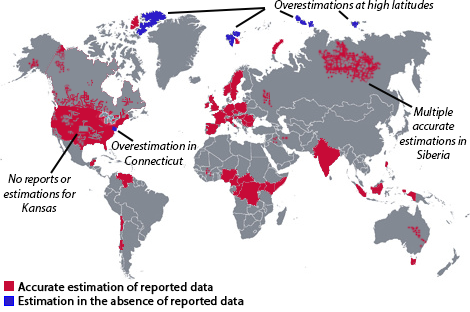}
\vspace{3mm} 
  \caption{\textbf{Estimations of parameter \textit{location}~\cite{map}.  Noteworthy events are indicated.}  Red and blue denote correct and erroneous estimations of death, respectively.  Erroneous estimations occurred in Connecticut and on small islands near the Arctic circle.  Surprising dense reports from Siberia were estimated correctly.}
\end{figure}
\begin{multicols}{2}
Other events within the prediction window seem reasonable.  The Woods occurs often, in agreement with our preconceived speculation that the Woods should occur often.  Already we have confirmed the 3:10:59 event in Norway~\cite{private1}, which impressively was mere seconds off.  Since 6am we have obtained off-the-books evidence of two deaths in some woods in Northeastern Texas~\cite{private2,private3}.  Both are near the predicted 4:50:22 event in Nacogdoches.  Currently we are working to quantify the likelihood that either of these Texas reports represents the Nacogdoches prediction, given the error covariance between state $x$ and location.  

Finally, the one disquieting prediction implicates the Woods on Tavira Island, Portugal.  There are no woods on Tavira Island, Portugal~\cite{taviraIslandPortugal}.  It is possible that, for some 
\end{multicols}
\begin{figure}[H]
\centering
  \includegraphics[width=\textwidth]{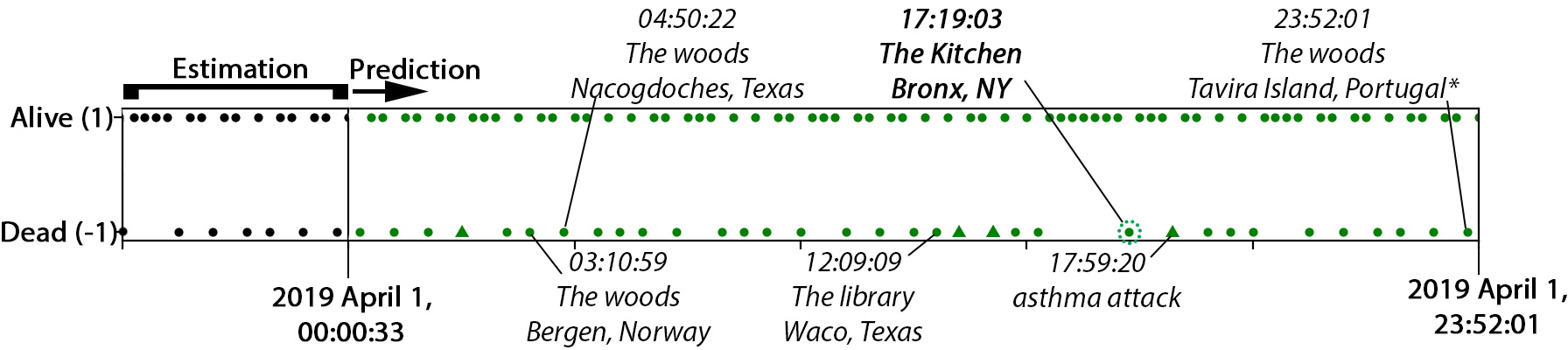}
  \caption{\textbf{Estimation and prediction window for the observed state variable $x$}, with details given for some events.  The prediction window begins at 00:00:33 on April 1, and the next local event at 17:19:03 is denoted in bold.  For Dead points, circles and triangles represent murders and asthma attacks, respectively.  *There are no woods on Tavira Island, Portugal; see text. }
\end{figure}
\begin{multicols}{2}
\noindent
local neighborhoods in the time series, location is a sloppy direction in state space.  In addition, this prediction occurs later into the day, and the rate of divergence of paths may be especially rapid in the direction of location.  Examining the model for specific locations of chaos, however, is beyond the scope of this paper.
\end{multicols}
\section{DISCUSSION}
\begin{multicols}{2}
\subsection{Who is/was Mr. Boddy?}
While the estimation window overall appears reasonable, there occurred a few peculiar events, including the erroneous estimation in Connecticut and the prediction in nonexistent woods on an island off Portugal.  We have acknowledged various assumptions throughout the modeling procedure, any of which may have introduced errors into the results.  In particular, the assumption of an over-simplistic model may be a problem.

We wrote Mr. Boddy's internal dynamics solely in terms of his binary alive-or-dead state and his IgE levels.  For future iterations of estimation we plan to flesh Mr. Boddy out a bit.  For example, it might be useful to consider what kind of a human being he is/was.  Aspects of his personality might yield insight into what he was up to at \textit{The Kitchen} in the Bronx, or in Siberia so frequently.  Or, for that matter, in a nine-room mansion for 70 years. 

\subsection{Quantum effects?}
We'd like to address an apparent paradox in case it has been noted by the reader.  On one hand, quantum considerations should not enter into the dynamical evolution of a typically-sized human being wandering about the planet at some typical rate of motion.  For this reason we imposed unitarity upon Mr. Boddy as a constraint in the cost function.  On the other hand, there has occurred a handful of instances at which Mr. Boddy was reported to be simultaneously alive and dead.  Now, the formulation of quantum mechanics indeed permits Mr. Boddy to exist as a superposition of Alive and Dead states - \textit{until he is observed}.  Here, we have allegations of simultaneous \textit{observations} of multiple state values.  In other words, these allegations belie quantum mechanics itself.  

This might feel unsettling, but for a caveat: the time resolution of reported deaths is one second at best.  Thus, the simultaneity of reports is questionable.  At any rate, to grapple with them is beyond the scope of this paper.

\subsection{The next murder: confirm or prevent?}
Currently we are en route to The Bronx on the D train, to await \textit{The Kitchen} murder, and we recognize an ethical dilemma: to attempt to prevent the murder?  If we intervene, we will impose an additional external force upon the dynamics, thereby altering the model.  That is, to prevent the murder would preclude an assessment of the model's predictive power.  Meanwhile, in principle, preventing the murder is the right thing to do~\cite{bible}.  We note, however, the qualifier \lq\lq in principle\rq\rq, because in Mr. Boddy's case murder does not appear to exert its characteristic distasteful influence.  Hm.  Well, there remain a few hours yet for soul-searching.  
\begin{figure}[H]
\centering
  \includegraphics[width=50mm]{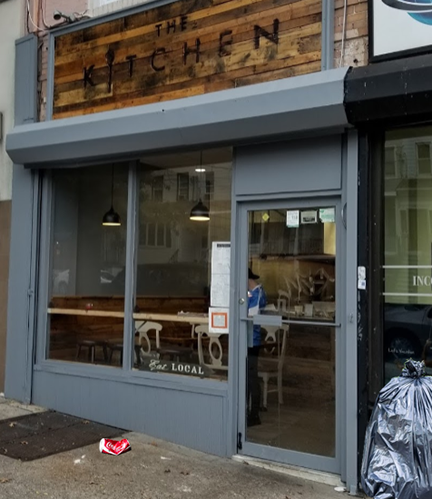}
  \caption{\textbf{Next local prediction: \textit{The Kitchen}, Bronx, NY.} }
\end{figure}
\end{multicols}

\bibliographystyle{unsrt}
\nocite{*}
\bibliography{refs}


 



\end{document}